\documentclass[11pt]{article}
\usepackage[preprint]{acl}
\ifdefined\ANONYMIZED
\makeatletter
\acl@anonymizetrue
\makeatother
\fi

\usepackage{times}
\usepackage{latexsym}
\usepackage[T1]{fontenc}
\usepackage[utf8]{inputenc}
\usepackage{microtype}
\usepackage{graphicx}
\usepackage{amsmath}
\usepackage{booktabs}
\usepackage{url}
\usepackage{tikz}
\let\origunderscore\_
\renewcommand{\_}{\origunderscore\allowbreak}
\usetikzlibrary{shapes.geometric,shapes.misc,arrows.meta,positioning,fit,backgrounds}

\title{FlinkEval: A Synthetic Corpus for Evaluating\\
Natural Language Interfaces to Streaming SQL}

\ifdefined\ANONYMIZED
\else
\fi

\ifdefined\ANONYMIZED
\author{}
\else
\author{Alfio Massimiliano Gliozzo \\
  IBM \\
  \texttt{gliozzo@us.ibm.com} \And
  Nandana Mihindukulasooriya \\
  IBM \\
  \texttt{nandana@ibm.com} \AND
  Gaetano Rossiello \\
  IBM \\
  \texttt{gaetano.rossiello@ibm.com} \And
  Erick Lee \\
  Confluent \\
  \texttt{erick@confluent.io} \AND
  Lisa Amini \\
  IBM \\
  \texttt{lisa.amini@us.ibm.com}}
\fi

\begin{document}
\maketitle

\begin{abstract}
Existing natural-language-to-SQL benchmarks primarily evaluate queries over
bounded relational data, leaving streaming semantics underexplored.
We introduce \textbf{FlinkEval}, an execution-grounded synthetic benchmark
comprising 3,903 question--SQL pairs across 45 industry domains and 45
semantic labels. We create diverse schemas and domain-plausible,
temporally consistent data through iterative generation
and validation. SQL patterns collected from the human-written NL2SQL pairs and
vendor documentation drive quota-balanced generation, and execution with a
repair loop on Confluent Cloud Apache Flink produces reference queries covering
windowing, temporal joins, pattern matching, nested types, and data manipulation. The corpus includes
paired schema-explicit and business-language requests, with annotations for
analysis by query pattern, complexity, and execution mode. A seven-level
Question Specification Ladder examines semantics-preserving reformulation
and underspecified requests augmented with oracle-derived clues. Baseline
evaluations combine execution outcomes with adjudicated LLM judgments,
revealing challenges in pattern composition and request specification.
\end{abstract}

% -------------------------------------------------------
\section{Introduction}
\label{sec:intro}

Natural-language interfaces to databases (NL2SQL) have advanced substantially
through benchmarks such as Spider \cite{yu-etal-2018-spider}, Spider~2.0
\cite{lei-etal-2024-spider2}, and BIRD \cite{li2023bird}. These benchmarks primarily address queries over bounded
relational data. In contrast, streaming engines such as Apache Flink operate
over evolving inputs whose interpretation depends on event time, watermarks,
temporal relationships, and changelog behaviour. Consequently, successful
translation requires more than selecting tables and columns: the generated
query must respect the target engine's temporal and operational semantics.

We study this problem in Confluent Cloud (CC), a managed platform built on
Apache Kafka and Apache Flink. Its SQL interface supports continuously
updated streaming queries, bounded snapshots of stream state, and background
data manipulation (DML) that writes results to Kafka topics. For example,
a syntactically valid pattern-matching query
may fail when its input produces updates rather than append-only events.
Such constraints motivate evaluation against an operational engine
(Appendix~\ref{sec:appendix:sql-examples}).

A second challenge concerns the specification of user intent. A business
request such as ``Which of my recent ad impressions were actually viewable?''
may leave the data source or the meaning of ``recent'' implicit. Some omitted
requirements may be recoverable from the available schema or domain
definitions; others may require clarification. This distinction matters for
evaluation: disagreement with a single reference query need not establish
that a prediction is incompatible with the request
\cite{bhaskar-etal-2023-ambiguity,sarwar-etal-2026-clarity,li-etal-2026-underspecification}.
Conversely, a plausible prediction does not demonstrate that the user's
intended semantics have been resolved.

To investigate both challenges, we introduce \textbf{FlinkEval}, a synthetic
benchmark grounded in a live CC Flink deployment. We generate candidate SQL
from domain schemas, retain queries that pass execution validation, and
construct corresponding natural-language requests. The accompanying
\emph{Question Specification Ladder} distinguishes identifier abstraction
and paraphrasing from potentially underspecified requests augmented with
oracle-derived clues. It supports comparisons across information conditions
without presuming that every omitted detail constitutes genuine ambiguity.

Our evaluation addresses two questions: how reliably do current LLMs generate
Flink SQL across execution modes and query patterns, and how does performance
depend on the specification provided by the request? Five-model comparisons
reveal substantial differences in execution reliability and performance on
queries combining multiple patterns. A separate specification study finds
similar accuracy under identifier abstraction and paraphrasing, but markedly
lower agreement with the gold SQL for unclarified requests. The sensitivity
of this result to the judging criterion motivates distinguishing reference
agreement from validated resolution of user intent.

Our contributions are threefold:
\begin{itemize}
  \item \textbf{A streaming NL2SQL corpus.} We synthesise 50 domain schemas
        and temporally consistent data through structural validation, live
        integrity checks, and plausibility assessment
        (Appendix~\ref{sec:appendix:datagen}), and generate 3,903 NL--SQL
        pairs across 45 domains and 45 semantic labels from 46 SQL patterns
        collected from human-written pairs and vendor documentation
        (Appendix~\ref{sec:corpus:patterns}). Every reference SQL executes on
        a live CC Flink cluster and returns a non-empty result
        (Appendix~\ref{sec:corpus:gen}).
  \item \textbf{A five-model evaluation.} We score execution outcomes with an
        adjudicated three-judge panel and characterise performance by query
        complexity, execution mode, and semantic pattern
        (Section~\ref{sec:exp1}).
  \item \textbf{The Question Specification Ladder.} We analyse seven
        specification conditions under ground-truth-anchored and face-value
        criteria (Section~\ref{sec:exp2}).
\end{itemize}

% -------------------------------------------------------
\section{Related Work}
\label{sec:related}

\paragraph{NL2SQL benchmarks.}
Spider \cite{yu-etal-2018-spider} established cross-domain NL2SQL evaluation
with 10,181 questions over 200 databases, and BIRD \cite{li2023bird} added
12,751 queries over large real-world databases with external knowledge.
FlinkEval complements these resources by evaluating temporal and operational
constraints specific to streaming SQL.

\paragraph{Ambiguity, underspecification, and clarification.}
AmbiQT \cite{bhaskar-etal-2023-ambiguity} constructs questions with two
plausible SQL interpretations arising from lexical or structural ambiguity.
CLARITY \cite{sarwar-etal-2026-clarity} distinguishes lexical and semantic
ambiguity from unanswerability and studies partially helpful clarification.
PLANPOOL \cite{li-etal-2026-underspecification} defines underspecification
through multiple plausible SQL semantics and separates execution correctness
from grounded resolution. FlinkEval draws on these distinctions to compare
linguistic reformulation with omitted semantic information
(Section~\ref{sec:task:complexity}).

\paragraph{Synthetic data for NL2SQL.}
\citet{wu-etal-2021-data} generate synthetic question--SQL pairs through
SQL-query generation and hierarchical SQL-to-question generation, improving
cross-domain text-to-SQL parsing. Complementary approaches use intermediate
representations and schema linking \cite{guo-etal-2019-towards}, or decompose
LLM-based SQL generation into reasoning and correction stages
\cite{pourreza2023dinsql}.
Our approach uses execution on a live cluster to validate synthetic reference
queries against the target engine's constraints.

\paragraph{LLM-as-judge.}
LLM-based evaluation offers a scalable alternative to manual assessment,
although judgments can depend on evaluator biases and calibration
\cite{zheng2023judging,liu2023evaluating}.
We apply this approach to streaming SQL through \emph{panel adjudication},
in which a third judge resolves disagreements between two primary judges.
Comparing ground-truth-anchored and face-value criteria further exposes
sensitivity to how the evaluation treats implicit requirements.

\paragraph{Streaming SQL.}
Apache Flink unifies stream and batch processing \cite{carbone2015apache}.
Its SQL interface provides window TVFs, watermarks, and temporal joins
\cite{confluent2026sql}, and Confluent Cloud adds dialect extensions that
FlinkEval targets.

% -------------------------------------------------------
\section{The NL2Flink-SQL Task}
\label{sec:task}

Given a natural-language question $q$ and a set of Flink table DDLs
$\mathcal{S}$, a system under evaluation produces a Flink SQL statement
$\hat{s}$.
FlinkEval provides a gold SQL $s^*$ and, for bounded queries, a reference
result set $r^*$ obtained by executing $s^*$ on a live CC Flink cluster.

% ---- 3.1 Question specification ----
\subsection{Question specification and underspecification}
\label{sec:task:complexity}

Requests vary in both linguistic form and how fully they specify intended
SQL semantics. We distinguish \emph{Contextually recoverable underspecification}, replacing
technical identifiers with everyday expressions without intentionally
altering query constraints, from \emph{Contextually unresolved underspecification}, omitting
or leaving implicit SQL-relevant information such as projections, sources,
joins, predicates, time windows, ranking, aggregation, or business metrics.

% Informed by ambiguity and clarification research
% \cite{bhaskar-etal-2023-ambiguity,sarwar-etal-2026-clarity,li-etal-2026-underspecification},
% we propose two analytical categories for FlinkEval.
% \emph{Contextually recoverable underspecification} leaves requirements implicit
% but sufficiently grounded in the permitted schema, documentation,
% organisational definitions, data, or conversation, without unsupported assumptions.
% \emph{Contextually unresolved underspecification} leaves multiple materially
% distinct choices plausible under that context, requiring clarification or an
% explicitly declared assumption to resolve the user's particular intent.
% We introduce these categories as FlinkEval terminology informed by the cited
% work. Their application is context-dependent: supplying a semantic layer or
% business definition can change whether an omitted requirement is recoverable.

% \emph{Ambiguity} requires more than one plausible semantic interpretation or
% SQL intent; shortness or schema abstraction alone does not establish it.
% \emph{Unanswerability} means the available data or schema cannot support the
% requested information, rather than merely leaving an available parameter,
% such as a time horizon, unspecified.

Table~\ref{tab:specification-ladder} defines our seven-level
\emph{Question Specification Ladder}, which operationalises variation in
request specification rather than a universal ordering of difficulty.
The table lists the levels from the most to the least specified request
(L0 to L2.1); within L2 the rows run from three appended clues (L2.4) down to
none (L2.1), the base request that the others extend.
L1.1 and L1.2 are generated to preserve semantics and are verified for
equivalence. L2.1 denotes a potentially underspecified generation condition
relative to a single gold SQL; it does not establish item-level ambiguity.
In the corpus, \texttt{nl\_input\_generated} stores the schema-explicit
source form (L0) and \texttt{nl\_question\_simple} the unclarified
business-language base form (L2.1).

\begin{table*}[t]
\centering\small
\begin{tabular}{@{}lp{4.5cm}p{9.5cm}@{}}
\toprule
\textbf{Level} & \textbf{Operational label} & \textbf{Specification condition} \\
\midrule
L0 & Schema-Explicit Specification & Fully specified NL with explicit technical schema identifiers, operations, predicates, values, and relevant parameters. \\[3pt]
L1.1 & Schema-Abstracted Specification & Replace table and column identifiers with natural-language equivalents, aiming to preserve complete intended SQL semantics. \\[3pt]
L1.2 & Semantics-Preserving Paraphrase & Reformulate in ordinary business language without deliberately removing necessary information. \\[3pt]
L2.4 & Clue-Augmented Request (+3) & Append the top three ranked SQL-relevant clues to the L2.1 base request. \\[3pt]
L2.3 & Clue-Augmented Request (+2) & Append the top two ranked clues to the same base request. \\[3pt]
L2.2 & Clue-Augmented Request (+1) & Append the top ranked clue to the same base request. \\[3pt]
L2.1 & Unclarified Underspecified Request & Business-language request before clues; omitted constraints may be recoverable from context or require clarification. \\
\bottomrule
\end{tabular}
\caption{Question Specification Ladder in FlinkEval, ranging from schema-explicit
requests to unclarified, potentially underspecified business-language requests;
rows run from most to least specified.
L2.2--L2.4 augment L2.1 with ranked oracle-derived clues. Results appear in
Section~\ref{sec:exp2}, Table~\ref{tab:exp2}.}
\label{tab:specification-ladder}
\end{table*}

Levels L2.1--L2.4 share the same base question and differ only in how many
\emph{clues} are appended.
% A clue is a ranked, atomic piece of SQL-relevant information extracted by an
% LLM from the schema-explicit source form and the ground-truth SQL.
% Formally, let $q_{2.1}$ be the unclarified base request and $c_1, c_2, c_3$ the ranked
% clues; then $q_{2.2} = q_{2.1} \| c_1$,
% $q_{2.3} = q_{2.1} \| c_1 \| c_2$,
% $q_{2.4} = q_{2.1} \| c_1 \| c_2 \| c_3$,
% where $\|$ denotes concatenation with a natural-language connector
% (``\emph{Hint:~\ldots}'').
% This nesting guarantees inclusion of earlier clues, not monotonically
% improving model performance on each item.
The LLM ranks clues by their estimated usefulness in resolving omitted
SQL-relevant details (typically sources, filters, aggregation keys, or time
intervals). %the ranking is not independently validated as an ambiguity reduction.
\subsection{Evaluation metrics}
\label{sec:task:metrics}

We report execution outcomes and LLM judgments, which capture complementary
aspects of correctness. 

\paragraph{Execution-error rate.}
Execution-error rate is the proportion of evaluated predictions with a
recorded error during SQL submission or result retrieval under the benchmark's
execution protocol (Appendix~\ref{sec:appendix:configuration}). It captures
operational failures, including parsing, validation, and observed runtime
errors, rather than agreement with reference results. For streaming queries,
the observation period is bounded; absence of an error does not establish
indefinite execution success or semantic correctness. 
We do not report execution accuracy (ExAcc): comparing evolving streaming
results requires aligned replay, defined observation horizons, and handling
of retractions (Appendix~\ref{sec:appendix:configuration}).

\paragraph{LLM-as-judge (judge-pass rate).}
Our primary metric is judge-pass rate, following the LLM-as-judge paradigm of
\citet{zheng2023judging} and G-Eval \cite{liu2023evaluating}.
A judge LLM receives the question, the DDLs, the gold SQL and its execution
output, and the predicted SQL and its execution output, returning a ternary
verdict: \textit{pass} (1.0), \textit{maybe} (0.5), or \textit{fail} (0.0).
Judge-pass rate is the fraction of \textit{pass} verdicts.

We use a \emph{three-judge panel}. Two primary judges score each prediction independently,
and a third model serves as adjudicator in case of disagreement. 
% in the toolkit scoring path, the draw uses
% the question and predicted SQL to identify the instance.
% The implemented combination rule assigns 1.0 to pass/pass and pass/maybe
% pairs, and 0.0 to fail/fail pairs. Other pairs, including maybe/maybe, and
% primary-judge errors invoke the adjudicator, which receives both verdicts
% and explanations. Thus adjudication frequency is not identical to the
% frequency of differing primary verdicts. 
% If adjudication fails, the mean of the available primary scores
% is retained and flagged as \texttt{adjudicator\_error\_fallback}.
Judge-pass rate counts only final scores equal to 1.0. We use
\emph{non-pass} for all scores below 1.0.

% \begin{table}[t]
% \centering\small
% \begin{tabular}{lp{4.8cm}}
% \toprule
% \textbf{Metric} & \textbf{Definition} \\
% \midrule
% Judge-pass\%  & \% questions with final score = 1.0 \\
% Exec error\%  & \% predictions with a recorded submission or retrieval error \\
% Pass \& ran\% & \% questions with score = 1.0 and no execution error \\
% \bottomrule
% \end{tabular}
% \caption{Reported evaluation metrics. Percentage denominators include all
% evaluated questions in the corresponding condition.}
% \label{tab:metrics}
% \end{table}

% -------------------------------------------------------
\section{FlinkEval Construction}
\label{sec:generation}

FlinkEval was built iteratively, with execution on a live CC Flink cluster as
the common acceptance criterion. Human expertise enters at three points:
subject-matter experts (SMEs) wrote and validated the initial NL2FlinkSQL
pairs from which most SQL patterns are derived, audited a sample of synthetic
pairs to identify systematic defects, and formulated the engine rules that
the generator enforces as guardrails and lint checks. Generation, execution
and labelling are otherwise automatic.
% This section describes the current release, v3; the model comparison in
% Section~\ref{sec:experiments} was run on the v3.1 test split.

% \subsection{Corpus releases}
% \label{sec:corpus:versions}
\subsection{Synthetic Generation of the Benchmark}
We start from 30 NL--SQL pairs written by SMEs; their statement mix follows
observed usage of the CC Flink SQL workspace and addresses everyday analytical
questions rather than dialect edge cases. From these pairs and vendor
documentation we derive a catalogue of 46 SQL patterns, extend every schema
with an append-only, watermarked stream table and a table with nested columns
(Appendix~\ref{sec:appendix:datagen}), and generate queries under a quota
schedule, SME-derived dialect guardrails, and a repair loop
(Appendix~\ref{sec:corpus:gen}). The guardrails encode the failure modes the
SME audit exposed: time-relative predicates on historic data, streaming
constructs applied to upsert tables, Top-N rankings without a unique
tie-breaker, and pattern matching or \texttt{OVER} windows over non-append
input.

\subsection{Benchmark statistics}
\label{sec:corpus}

Table~\ref{tab:corpus} summarises the benchmark. It contains 3,903 NL--SQL
pairs over 50 schemas and 404 table DDLs in 45 industry domains, between 57 and
89 pairs per schema; four domains (hospital, insurance, retail, sports) are
covered by two or three schemas. Each pair has two generated requests, the L0
schema-explicit form (49.7 words on average) and the L2.1 business-language
form (24.1 words), and five further ladder variants (L1.1--L2.4), for 27,321
natural-language questions in total. All five variants pass LLM verification for
3,492 pairs (89.5\%); the remaining 411 pairs keep at least one variant that
failed verification, mostly L1.1 identifier abstraction (374 pairs).

\begin{table}[t]
\centering\small\setlength{\tabcolsep}{4pt}
\begin{tabular}{@{}lrr@{}}
\toprule
\multicolumn{3}{@{}l}{\emph{Scale}} \\
NL--SQL pairs                    & \multicolumn{2}{r}{3,903} \\
Domains / schemas / tables       & \multicolumn{2}{r}{45 / 50 / 404} \\
Observed patterns / groups       & \multicolumn{2}{r}{45 / 14} \\
Generated questions (7 levels)   & \multicolumn{2}{r}{27,321} \\
\midrule
\multicolumn{3}{@{}l}{\emph{Complexity tier}} \\
Tier 1 (single pattern)          & 1,659 & 42.5\% \\
Tier 2 (two patterns)            & 1,342 & 34.4\% \\
Tier 3 (three or more)           &   902 & 23.1\% \\
\midrule
\multicolumn{3}{@{}l}{\emph{Execution mode}} \\
Snapshot                         & 1,677 & 43.0\% \\
Streaming                        & 1,828 & 46.8\% \\
Background (DML)                 &   398 & 10.2\% \\
\bottomrule
\end{tabular}
\caption{The FlinkEval corpus. Generated questions: two stored requests
(L0, L2.1) and five ladder variants per pair. Per-group pair counts appear
with the pattern catalogue (Table~\ref{tab:pattern-groups}).}
\label{tab:corpus}
\end{table}

Complexity is assigned by pattern count; 57.5\% of the pairs combine two or more
patterns. Snapshot and streaming distinguish bounded from ongoing result
retrieval; background identifies DML statements that write to sinks. These are
operational categories rather than disjoint SQL pattern classes. Every group is
covered, and 13 of the 14 groups contain all of their patterns
(Table~\ref{tab:pattern-groups}).
%Filtering and conditional logic (G7) is the
%largest group because its constructs are mostly added as secondaries: 854 of
%its 1,150 pairs target a pattern from another group. 

% Every reference was accepted under the protocol of Section~\ref{sec:corpus:gen},
% and all 3,505 snapshot and streaming references have a stored, non-empty result
% capture (up to 200 rows; 1,156 captures are truncated). Of the 398 background
% statements, the 50 \texttt{CREATE TABLE \ldots AS} statements store the
% result of their inner query; the remaining 348 statement sets and Faker table
% definitions have no result capture and are scored by the judge alone. 

Each pair
contains the L0 and L2.1 requests, the reference SQL, semantic labels, the
planned and observed execution mode, the number of repairs, and the target
pattern; the ladder variants are stored in sidecar files
(Appendix~\ref{sec:appendix:schema}).

% \paragraph{Reference-result stability.}
% Bounded population runs terminate, but subsequent writes or repopulation
% can change topic contents. Because v3 excludes clock-relative predicates,
% references depend only on the data snapshot. Captures from streaming queries are
% bounded observations rather than complete results of unbounded computations,
% and should be interpreted with their snapshot, capture time and truncation
% status (Section~\ref{sec:task:metrics}).

% -------------------------------------------------------
\section{Experiments}
\label{sec:experiments}

We investigate two research questions: \textbf{RQ1}, how generation quality
varies across models and streaming SQL constructs; and \textbf{RQ2}, how
request specification affects reference agreement and judged acceptability.
The studies share an execution and adjudication protocol but use different
question samples. The model comparison (Section~\ref{sec:exp1}) was run on
the held-out test split of release v3.1, and so was the Ladder study
(Section~\ref{sec:exp2}).
%Both were run on the v2 release (3,940 pairs, 31 labels;
%Section~\ref{sec:corpus:versions}), whose label groups are a subset of those
%in Table~\ref{tab:pattern-groups}.

\paragraph{Shared evaluation protocol.}
We evaluate simple baselines rather than optimising them. A single LLM
receives a two-message prompt (system and user) with the natural-language
question and the domain table DDLs; no few-shot examples, retrieval or repair
are used. Every prediction is executed on the live Confluent Cloud cluster and
scored by the adjudicated three-judge panel of Section~\ref{sec:task:metrics}
(Claude Opus~5.5, GPT-6-Astra, Gemini~3.6 Flash, seed~0).

%The ground-truth-anchored criterion uses the gold SQL to interpret
%the request rather than as a strict string or structure to match. It
%compares relational and temporal semantics (filters, join cardinality,
%grouping, time source and range) even when the two result sets happen to
%coincide.
%We report judge-pass rate with Wilson 95\% confidence intervals and compare
% models on the same questions with an exact McNemar test.
% Appendix~\ref{sec:appendix:configuration} records model identifiers and the
% evaluation runner's execution procedure.

Both studies provide DDLs for all tables in the question's domain as schema
context (about 1.6k prompt tokens on average). The model comparison uses L0
requests; the specification study varies the level $L$ while holding the
domain context fixed, enabling within-question comparisons of the information
supplied by the request.

% ---------------------------------------------------------
\subsection{Model comparison on schema-explicit requests}
\label{sec:exp1}

\paragraph{Experimental setup.}
We evaluate every model on the same 666 questions of the held-out test split
of release v3.1 (781 questions), which cover all 45 domains (11--42 per
domain). We exclude the 115 test questions whose reference SQL uses a
\texttt{TUMBLE} or \texttt{HOP} group window: submitted as batch queries
they trigger a known Confluent Cloud Flink defect (SQL-5248) that leaves the
statement pending and pages the service's on-call engineers, so they could not
be run safely. The exclusion depends only on the reference query and is
therefore identical for every model.
Each of five models receives the L0 question and the domain-specific table DDLs,
yielding 3{,}330 statements that are submitted for cluster execution.
The models are GPT-6-Astra, GPT-5.6-Sol, Claude Opus~5, and Gemini~3.6
Flash as frontier models, and GPT-OSS-120B as an open-weight model
(served on watsonx.ai).
The judging procedure completed for all 3{,}330 statements.

\paragraph{Overall comparison.}
The largest performance difference in Table~\ref{tab:exp1-slices} separates the four
frontier models (65--80\% judge-pass rate) from GPT-OSS-120B (35.0\%;
$p<10^{-45}$ against every frontier model, exact McNemar). Claude Opus~5 and
GPT-6-Astra are the two strongest models and cannot be distinguished
($p=0.93$); each is significantly above GPT-5.6-Sol ($p<10^{-6}$) and
Gemini~3.6 Flash ($p<10^{-13}$), and GPT-5.6-Sol is above Gemini~3.6 Flash
($p=0.003$). Tests are unadjusted. Discordant-pair counts are in
Appendix~\ref{sec:appendix:experiments}.

\begin{table*}[t]
\centering\small
\begin{tabular}{lrrrrr}
\toprule
\textbf{Slice} & \textbf{GPT-6} & \textbf{Claude}
  & \textbf{GPT-5.6} & \textbf{Gemini} & \textbf{OSS-120B} \\
\midrule
All questions (666) & 79.3 & \textbf{79.6} & 70.0 & 65.0 & 35.0 \\
95\% CI (Wilson) & 76.0--82.2 & 76.4--82.5 & 66.4--73.3 & 61.3--68.5 & 31.5--38.7 \\
Execution errors & 21.5 & \textbf{18.8} & 25.5 & 30.3 & 49.4 \\
Pass \& ran & \textbf{68.3} & \textbf{68.3} & 59.9 & 56.0 & 34.8 \\
\midrule
\multicolumn{6}{l}{\emph{Complexity tier}} \\
Tier 1 (single pattern, 336) & 83.3 & \textbf{84.8} & 75.0 & 69.6 & 39.6 \\
Tier 2 (two patterns, 230)   & 79.1 & \textbf{79.6} & 70.0 & 65.2 & 32.6 \\
Tier 3 (three or more, 100)  & \textbf{66.0} & 62.0 & 53.0 & 49.0 & 25.0 \\
\midrule
\multicolumn{6}{l}{\emph{Execution mode}} \\
Snapshot (216)   & 83.3 & \textbf{84.3} & 75.0 & 69.4 & 44.4 \\
Streaming (369)  & \textbf{74.8} & 74.0 & 64.2 & 60.4 & 37.1 \\
Background (81)  & 88.9 & \textbf{92.6} & 82.7 & 74.1 & 0.0 \\
\bottomrule
\end{tabular}
\caption{Model comparison on the 666 test-split questions (L0 question, domain
DDL context, adjudicated ground-truth-anchored judge panel). Rows 1--2: judge-pass
rate (\%) and Wilson 95\% CI. Execution errors: share of statements that fail to
execute (all 81 background statements fail for every model, see text). Pass \&
ran: share of questions with score 1.0 and no execution error. Slice sizes in
parentheses; slices are not weighted to reproduce the corpus composition and
snapshot queries exclude the 115 \texttt{TUMBLE}/\texttt{HOP} questions
(Section~\ref{sec:exp1}). Best value per row in bold; pattern groups are in
Appendix~\ref{sec:appendix:experiments} (Table~\ref{tab:exp1-pattern-groups}).}
\label{tab:exp1-slices}
\label{tab:exp1}
\end{table*}

% \paragraph{Complexity tier.}
Table~\ref{tab:exp1-slices} shows lower accuracy on tier~3 than on tier~1
for every model, suggesting that pattern composition remains challenging.
GPT-6-Astra scores 83.3 / 79.1 / 66.0\% on tiers 1 / 2 / 3; the decline is
comparable or larger for Claude Opus~5 (84.8 / 79.6 / 62.0\%) and Gemini~3.6
Flash (69.6 / 65.2 / 49.0\%), and the tier-3 estimates remain uncertain
($n=100$).
% Aggregate rankings obscure this variation: Claude Opus~5 has the highest
% tier-1 accuracy but the lowest tier-3 accuracy among frontier models.
% Only 1 of GPT-6-Astra's 36 tier-3 statements fails execution, indicating that
% its lower pass rate primarily reflects semantic disagreement rather than
% execution failure. The tier-3 estimates nevertheless have considerable
% uncertainty ($n=36$; Wilson 95\% CI 53--82\% for GPT-6-Astra).

% \paragraph{Execution mode.}
Every model is less accurate on streaming queries than on snapshot ones, by
7--11 points.
All 81 background (DML) statements in the sample fail to execute for every
model: they write to sink tables that do not exist in the test environment,
or return no result set.
% Reference-query validation establishes compatibility at corpus-construction
% time; it does not establish that every prediction's required sink is present
% at evaluation time. These failures remain in the execution-error denominator.
% The four frontier models nevertheless receive \textit{pass} judgments on all
% 13 statements, whereas GPT-OSS-120B receives none.
The background-mode pass rates (74--93\% for the frontier models) therefore
reflect judged SQL quality rather than successful execution; the same
qualification applies to part of the ``statements and set operations'' group.
GPT-OSS-120B receives no credit on any background statement: none of its 81
answers uses an \texttt{EXECUTE STATEMENT SET} block, and the judges treat
several standalone \texttt{INSERT} statements as invalid for a single job.

% ---------------------------------------------------------
\subsection{Question Specification Ladder}
\label{sec:exp2}

To address RQ2, we compare performance across the ladder under two criteria:
agreement with a single gold SQL, and acceptability under plausible readings of
the presented request.
% This
%comparison measures sensitivity to the judging criterion; it does not
%independently validate alternative interpretations.

Each question is presented at the seven specification levels defined in
Section~\ref{sec:task:complexity} (Table~\ref{tab:specification-ladder}).
% The L2 base form is produced through broad
% business-language paraphrasing, not controlled deletion of individual
% semantic constraints.
% GPT-5.6-Sol (\texttt{azure/gpt-5.6-sol}) generates and verifies the intermediate
% variants.
% L1 generation receives the source request, with referenced table and column
% names additionally supplied for L1.1. Clue extraction receives the source
% and base requests, gold SQL, and referenced identifiers. L2 variants are
% constructed incrementally from the base request and ranked clues.
% The verifier checks L1 information preservation against L0 and L2 content
% against the base request and allowed clues. A failed variant is regenerated
% once using verifier feedback. The SQL-generation model, domain DDL context,
% prompt template, and execution procedure are held fixed across levels.

\paragraph{Experimental setup.}
We use the held-out test split of release v3.1 and keep the 600 questions, from
all 45 domains, for which all five generated variants passed verification and
whose reference SQL does not use a \texttt{TUMBLE} or \texttt{HOP} window
(Section~\ref{sec:exp1}); of the 781 test questions, 115 are excluded for the
window defect and a further 66 lack a complete set of verified variants.
% All five generated variants must be present and verified for a question to
% enter any condition. Of 257 candidate questions (67 / 98 / 92), 54 are
% excluded: 51 fail L1.1 source-table preservation checks (10 / 16 / 25), and
% three have fewer than three extracted clues (0 / 2 / 1). Retention is thus
% 79.0\%. The source-table checks sometimes reject shortened identifiers as
% changes of table identity, making selection sensitive to the verifier's
% interpretation of identifier abstraction.
GPT-6-Astra generates SQL for each question at all seven levels, yielding
4{,}200 statements. The L0 statements are those of the model comparison.
The same adjudicated three-judge panel evaluates every statement twice.
Twelve generated statements (1 / 3 / 1 / 7 at L2.4 / L2.3 / L2.2 / L2.1) use a
\texttt{TUMBLE} or \texttt{HOP} window although the reference does not; they
were not submitted to the cluster and are judged without an execution result.
The ground-truth-anchored criterion uses the gold SQL to interpret the
request; the ``face-value'' criterion treats it as one acceptable reading.
%Both evaluations receive the question at the tested level and share the
%same execution outcome.
We report pass rates with Wilson 95\% confidence intervals and compare
levels on the same questions with a paired exact McNemar test.

\begin{table*}[!tb]
\centering\footnotesize\setlength{\tabcolsep}{3pt}
\begin{tabular}{@{}lp{4.4cm}rrrrrr@{}}
\toprule
\textbf{Level} & \textbf{Specification condition} & \textbf{GT\%} & \textbf{95\% CI}
  & \textbf{Face\%} & \textbf{95\% CI} & \textbf{Gap}
  & \textbf{Err.\%} \\
\midrule
L0   & Schema-Explicit            & \textbf{79.5} & 76.1--82.5 & \textbf{79.7} & 76.3--82.7 &  0.2 & 21.3 \\
L1.1 & Schema-Abstracted          & 78.0 & 74.5--81.1 & 79.2 & 75.7--82.2 &  1.2 & 22.2 \\
L1.2 & Paraphrase                 & 78.2 & 74.7--81.3 & 78.2 & 74.7--81.3 &  0.0 & 22.2 \\
L2.4 & Clue-Augmented (+3)        & 51.3 & 47.3--55.3 & 73.7 & 70.0--77.0 & 22.3 & 22.2 \\
L2.3 & Clue-Augmented (+2)        & 46.3 & 42.4--50.3 & 74.2 & 70.5--77.5 & 27.8 & 21.7 \\
L2.2 & Clue-Augmented (+1)        & 40.2 & 36.3--44.1 & 73.2 & 69.5--76.6 & 33.0 & 21.3 \\
L2.1 & Unclarified                & 27.0 & 23.6--30.7 & 74.8 & 71.2--78.1 & 47.8 & 21.2 \\
\bottomrule
\end{tabular}
\caption{GPT-6-Astra on the Question Specification Ladder ($n=600$ test-split
questions per level, release v3.1; adjudicated judge panel). Face: judge
(face-value prompt v3) treating the gold SQL as one acceptable reading; gap:
face minus GT (pp); err.: statements that fail to execute, which includes the 75
background-mode statements (12.5\% of questions): all of them fail at L0--L1.2
and 53--60 at L2.1--L2.4. Clues are ranked oracle additions, not clarification
turns. Rows follow the order of Table~\ref{tab:specification-ladder}.}
\label{tab:exp2}
\end{table*}

\paragraph{Effect of request specification.}
Table~\ref{tab:exp2} shows similar accuracy for the three conditions intended
to preserve complete specification.
Identifier abstraction and paraphrasing (L1.1, L1.2) do not significantly change the pass
rate (L0 79.5\%, L1.1 78.0\%, L1.2 78.2\%; paired exact McNemar $p\ge0.30$).
The unclarified base condition (L2.1) instead yields 27.0\%, 52.5 points
below L0 (337 questions change from pass to non-pass and 22 in the
opposite direction, $p<10^{-72}$). Oracle clues partially recover reference
agreement: one clue raises the pass rate by 13.2 points
(27.0 $\to$ 40.2\%, $p<10^{-8}$), a second by a further 6.1 points
(46.3\%, $p=0.002$) and the third by 5.0 points (51.3\%, $p=0.011$).
The first clue closes about 26\% of the L1.2--L2.1 gap; with three clues the
rate is still 26.9 points below L1.2 ($p<10^{-33}$).
%Aggregate GT pass rates increase with appended clues;
%this does not guarantee improvement on each item, and face-value rates are
%not monotonic.
% At L2.1 the three domains are 38.6\% (education), 26.2\% (insurance) and
% 37.9\% (retail).
% Execution-error rates show no corresponding monotonic pattern (12.8--17.7\%).
% The additional non-pass judgments at lower specification levels primarily
% concern executable statements whose semantics differ from the reference.

This is consistent with an information-specification bottleneck, although
L2 changes wording and omits several details at once, which prevents
attribution to individual omissions.

\paragraph{Sensitivity to the evaluation criterion.}
Face-value judgments vary less across conditions: 78.2--79.7\% for L0--L1.2,
73.2--74.8\% for L2.4--L2.1, a drop of 4.5 points between L1.2 and L2.4
($p=0.002$) and no significant change among the four L2 levels (for example
L2.1 vs.\ L2.2, $p=0.35$). The two criteria give similar aggregate rates
for L0--L1.2 (gap 0.0--1.2 points, although at L0 59 questions pass only under
face-value and 58 only under ground truth) and diverge to 22.3, 27.8, 33.0 and
47.8 points for L2.4 to L2.1; at L2.1, 303 questions pass under face-value only
against 16 under the ground-truth criterion only.
% Its pass rate is 6--7 points above the GT judge for L0--L1.2, but 27--30
% points above for L2.4--L2.3 and 41--42 points above for L2.2--L2.1; the
% difference is significant at every level (e.g.\ 86 questions pass under
% face-value only against 1 under GT only at L2.1).
Reference agreement and judged plausibility thus diverge as explicit
constraints are removed.
% Human validation would be required
% to establish which additional predictions constitute acceptable alternative
% readings rather than evaluator errors. These rates therefore characterise
% two evaluation criteria, rather than bounds on semantic correctness or the
% prevalence of genuine ambiguity.

% \paragraph{Scope of interpretation.}
% The specification study uses one model, one run, and three domains, with
% LLM-generated and LLM-verified question variants.
% Its L0 pass rate (81.8\%) and the model-comparison result (85.3\%) are
% estimated on different question sets.
% The complete-variant filter further restricts the population to requests
% accepted by the same model used for variant generation.

% -------------------------------------------------------
\clearpage % body = pages 1--6; Conclusion and Limitations start on page 7
\section{Conclusion}
\label{sec:conclusion}

We introduced FlinkEval, an execution-grounded benchmark for natural-language
interfaces to streaming Flink SQL. Its cross-domain corpus enables evaluation
of temporal and operational requirements that are underrepresented in
conventional NL2SQL benchmarks. The five-model comparison identifies two
persistent challenges: generating executable statements for the target
engine and preserving intended semantics when multiple query patterns are
combined. 

On the Ladder, unclarified requests lose most reference agreement, oracle
clues recover only part of it, and the two judging criteria diverge.

\section*{Limitations}

\paragraph{Corpus and deployment scope.}
FlinkEval is grounded in a single Confluent Cloud cluster and reflects the
schema and data patterns of that deployment.
Generalisability to other Flink environments (e.g., Apache Flink on-premise,
other managed Flink services) requires additional validation.

\paragraph{Specification validity.}
The unclarified base requests (L2.1) and oracle clues (L2.2--L2.4) are LLM-generated and
not human-authored, which may introduce systematic biases toward phrasings the
generator itself favours.
Broad paraphrasing confounds linguistic changes with semantic omissions, and
the ladder lacks per-omission recoverability annotations. The single-gold
reference need not exhaust plausible intents, and L1 preservation is checked by
an LLM rather than an independent human study. Oracle-ranked clues do not
measure deployable retrieval or interactive clarification.

\paragraph{Evaluation reliability.}
The result-stability issues in Section~\ref{sec:task:metrics} complicate
execution-based comparison, while LLM judgments introduce evaluator uncertainty.
The panel invokes adjudication on a non-negligible share of predictions
(Section~\ref{sec:exp1}), and the evaluator pool overlaps with the generator
families; face-value judgments are not calibrated against human annotations.
The five-model comparison uses 666 questions of the v3.1 test split and a
single run per model; the 115 test questions that use \texttt{TUMBLE} or
\texttt{HOP} windows are excluded because of a Confluent Cloud defect
(SQL-5248), so batch tumbling and hopping window aggregates are not measured
in either study. The Ladder study uses a single model (GPT-6-Astra) and a
single run; full-corpus evaluation is available only for GPT-6-Astra on v2. These choices limit conclusions about small model
differences and run-to-run variability.

% -------------------------------------------------------
\bibliography{custom}

% -------------------------------------------------------
\clearpage
\appendix
% ===================== Appendix A =====================

\section{Data Generation: Schemas, Population, and Integrity}
\label{sec:appendix:datagen}

This appendix describes the construction of the database contents used by
the cluster-grounded pipeline in Section~\ref{sec:corpus:gen}.
% \emph{Confluent Synthetic Datagen} represents domain specifications and
% executable population workflows as machine-readable YAML documents.

\subsection{Schema design and coverage}

The source portfolio comprises 50 domain models across 45 business domains.
Multiple models can represent different aspects of the same domain. The
portfolio comprises 404 table DDLs: each model contributes an event stream and
a nested-type table in addition to its base tables.
Its ten structural families are normalised entity networks, header--detail,
many-to-many associations, hierarchies and graphs, supertype--subtype,
star, snowflake, fact constellations, temporal/event history, and
flexible or semi-structured data.
The sixteen targeted secondary features vary cardinality, optionality, key design,
relationship roles, join paths, attributed associations, join depth,
parallel child collections, grain and aggregation, temporal relationships,
tenant-scoped identity, connectivity, denormalisation, recursive topology,
subtype constraints, and flexible value shape.
These describe the source schemas; the SQL-pattern taxonomy in
Table~\ref{tab:pattern-groups} describes the generated queries.

For each target schema, an LLM receives the schema-design specification,
the domain-specification format, summaries of the other schemas, and Flink
capability constraints.
It specifies each table's row grain, columns, primary and unique keys,
relationships, value rules, and append/upsert semantics, together with a
real-time component, concrete data scenarios, and query intents.
A deterministic validator checks structural consistency, including key
references, compatible column types, tenant-scoped joins, and watermark
columns.
An LLM evaluator assesses whether the generated schema instantiates the
intended structural family and features; both assessments inform revision.
Accepted specifications are compiled into Flink DDL, including primary keys
declared \texttt{NOT ENFORCED} and event-time watermarks.

\subsection{Population workflows}

The workflow-generation model receives the domain specification, generated DDL,
an authoring guide with known Flink constraints, and a reference workflow.
It produces dependency-ordered steps using three population strategies:
\begin{enumerate}
  \item \textbf{Seed:} small reference and dimension tables receive fixed,
        domain-coherent rows with stable identifiers.
  \item \textbf{Faker:} bounded, append-only Flink Faker sources generate raw
        identifiers, categorical choices, numbers, and time offsets.
        An \texttt{INSERT\ldots SELECT} computes the target columns,
        drawing foreign keys from existing parents and deriving dependent
        values together.
  \item \textbf{Derived:} child records, events, versions, and aggregate
        snapshots are computed from populated tables in snapshot mode.
        Parent-key-based identifiers and sequence numbers preserve record
        linkage and ordering.
\end{enumerate}
Explicitly constructed records instantiate boundary conditions, such as an
open-ended version, an empty nested collection, or events on opposite sides
of a window boundary.
Restricting these conditions to designated records preserves variation in
the remaining population. Workflow execution follows the dependency order,
parallelises independent steps, and records statement and assertion outcomes.
Bounded generation ensures termination while retaining event-time
and changelog structure in the resulting Kafka-backed tables.

\subsection{Integrity validation and plausibility assessment}

A static linter checks that workflow DDL matches the specification, sources
are bounded, dependencies are valid, and every model table is populated and
checked.
Live SQL assertions then test row counts, key uniqueness, foreign-key
resolution (including composite keys), allowed values, ranges, and
domain-specific constraints such as non-overlapping intervals and subtype
consistency.
Each assertion returns a scalar compared with an expected value or bounds;
statistical properties can be reported as warnings.
Failed statements and assertions provide feedback for workflow revision.

An additional LLM judge reviews the seed values, Faker settings, and
derivation SQL for value plausibility, within-row and cross-table coherence,
codes and formats, locale coherence, and distributional and temporal plausibility.
Each finding identifies the affected step and cites supporting SQL, which is
verified against the workflow before revision.
A repair is accepted only after linting, a fresh live population run, and
re-review yield fewer errors, or the same errors with fewer warnings.
The repair mechanism preserves DDL and rejects deletion of existing checks,
error-to-warning downgrades, and removal of expectations.

For example, the charging workflow initially assigned 1.25\,kWh to a
two-second interval on a 150\,kW connector, implying 2,250\,kW.
The repair retained the temporal boundary case but reduced its energy to
approximately 0.0028\,kWh and added an assertion comparing interval power
with connector capacity.
A documented 50-schema generation run passed all 4,207 live integrity checks;
iterative plausibility assessment reduced error findings from 52 to four, leaving
46 schemas without error-level findings.
These are data-generation checks, complementary to the benchmark-query
execution filter in Section~\ref{sec:corpus:gen}.
Realism judgments remain model-based: the same model reviews its own
repairs, and residual temporal and distributional regularities can remain.

\subsection{Metadata and regeneration}

The population procedure attaches descriptive \texttt{model.*} metadata to Schema
Registry subjects, recording table grains, keys, relationships, measures,
and constraints alongside the physical schemas.
These properties document the domain model; integrity is checked by the
population workflow's assertions.
Versioned specifications, workflows, and generation logs support rebuilding
and inspecting the environment.
Fresh Faker draws and run-relative timestamps can change individual rows
across population runs, so exact data reuse requires preserving the populated
topics or copying a captured dataset using schema-aware export and import,
which preserve the record contents while remapping schema identifiers to
the destination environment.

% ===================== Appendix B =====================

\section{Corpus Construction: Patterns, Generation, and Labels}
\label{sec:appendix:patterns}

\subsection{Pattern acquisition}

\subsubsection{SQL pattern taxonomy}
\label{sec:corpus:patterns}

We derive SQL query patterns from the human-validated seed corpus and
supplement them with patterns from the
official \href{https://docs.confluent.io/cloud/current/flink/overview.html}{Confluent Cloud Flink SQL documentation} and \href{https://docs.confluent.io/cloud/current/flink/how-to-guides/overview.html}{how-to guides}.
The resulting taxonomy comprises 46 semantic patterns organised into
14 query groups (Table~\ref{tab:pattern-groups}).
These groups organise candidate generation. The retained corpus contains
45 observed labels, which the experiments aggregate into nine reporting
groups (Appendix~\ref{sec:appendix:groups}); generation groups and evaluation
groups therefore serve different purposes.

\begin{table}[t]
\centering\small\setlength{\tabcolsep}{4pt}
\begin{tabular}{@{}lrr@{}}
\toprule
\textbf{Source} & \textbf{Patterns} & \textbf{Assign.} \\
\midrule
SME-written seed pairs         & 26 & 4{,}617 \\
Long-running statement examples &  5 &   782 \\
Flink how-to guides            &  6 &   236 \\
Coverage expansion             &  9 & 1{,}773 \\
\midrule
Total                          & 46 & 7{,}408 \\
\bottomrule
\end{tabular}
\caption{Sources of the 46 SQL patterns; Assign.\ counts label
assignments (a pair can carry several labels).}
\label{tab:pattern-sources}
\end{table}

\paragraph{Seed-derived patterns.}
We first abstracted the reference SQL of the SME-written seed pairs into reusable
constructs. Each pair contributes the construct named in its rationale
(\texttt{temporal\_join}, \texttt{session\_window\_tvf},
\texttt{streaming\_topn\_rownumber}, \ldots), and multi-construct pairs
contribute several; for instance, the consecutive-orders fraud check
contributes \texttt{lag\_over\_window} and \texttt{timestampdiff}, and its
alternative formulation \texttt{match\_recognize}. Each pattern records the
identifiers of the seed pairs it was derived from, which preserves the link
between generated queries and the human-written examples: the entry in
Table~\ref{tab:corpus-example} carries \texttt{suite\_ids} 4, 18 and 25, the seed
pairs that define \texttt{temporal\_join}. This pass yielded
26 patterns, still the bulk of the catalogue (62\% of label assignments). Five seed pairs, all plain
\textsc{create table} or pass-through \textsc{insert} statements, define no
query pattern.

\paragraph{Documentation-derived patterns.}
We then added patterns from Confluent's material: five from a collection of
long-running statement examples (two- and three-table join aggregations,
self-join statistics, and unbounded inner and left joins), and six from
the CC Flink how-to guides
(\href{https://docs.confluent.io/cloud/current/flink/how-to-guides/overview.html}{online}),
including multi-stream \texttt{UNION ALL} with \texttt{LAST\_VALUE}, field
masking with \texttt{REGEXP\_REPLACE}, state-TTL and snapshot-mode settings,
and Faker-backed table definitions. SQL examples from both sources were
also executed on the cluster; the 16 how-to examples that run were retained as
reference material. The dialect rules in the generator prompt draw on the CC
Flink SQL reference \cite{confluent2026sql} and on the dialect gaps observed
in the seed pairs.

\paragraph{Coverage expansion.}
An SME audit of a pilot sample showed that common analytical constructs were missing entirely.
The catalogue adds nine patterns: \texttt{CUMULATE} windows, \texttt{CASE WHEN}
derivations, \texttt{IN}/\texttt{EXISTS} subqueries,
\texttt{COALESCE}/\texttt{NULLIF} handling, conditional aggregation with
\texttt{FILTER} and \texttt{HAVING}, \texttt{CROSS JOIN UNNEST}, \texttt{ROW}
field access, array and map functions, and JSON structure functions
(\texttt{JSON\_QUERY}, \texttt{JSON\_EXISTS}). These are grouped under a new
group G15 (complex and nested types) or existing groups. The nested and
JSON patterns are reachable only through the extension tables described in
Appendix~\ref{sec:appendix:datagen}.

\paragraph{Resulting taxonomy.}
The catalogue contains 46 patterns in 14 groups (Table~\ref{tab:pattern-groups}).
One pattern (\texttt{time\_series\_math}) is not generated, because it requires
synthetic signal generation, which yields no stable reference result. The detectors were revised after generation (e.g.,
interval joins with an offset lower bound, and unbounded \texttt{OVER} windows in
their \texttt{BETWEEN} form), and all labels were recomputed from the stored SQL.

\begin{table*}[t]
\centering\footnotesize\setlength{\tabcolsep}{5pt}
\begin{tabular}{@{}lp{3.2cm}p{9.6cm}@{}}
\toprule
\textbf{Group} & \textbf{Name} & \textbf{Patterns} \\
\midrule
G1  & Windowed aggregation   & \texttt{tumble\_tvf}, \texttt{hop\_tvf}, \texttt{session\_window\_tvf}, \texttt{windowed\_count\_having}, \texttt{cumulate\_tvf}$^{*}$ \\
G2  & Window join            & \texttt{window\_join} \\
G3  & OVER windows           & \texttt{over\_rows\_preceding}, \texttt{over\_range\_\allowbreak interval\_preceding}, \texttt{over\_window\_unbounded}, \texttt{lag\_over\_window}, \texttt{lead\_over\_window}, \texttt{avg\_over\_partition} \\
G4  & Dedup / Top-N          & \texttt{dedup\_rownumber\_eq1}, \texttt{streaming\_topn\_rownumber} \\
G5  & Stream--stream joins   & \texttt{stream\_stream\_interval\_join}, \texttt{unbounded\_inner\_join}, \texttt{unbounded\_left\_join} \\
G6  & Temporal / enrichment  & \texttt{temporal\_join}, \texttt{multi\_join\_chain}, \texttt{two\_table\_join\_agg}, \texttt{self\_join\_aggregation} \\
G7  & Filter / conditional   & \texttt{filter\_with\_interval\_bound}, \texttt{null\_empty\_filter}, \texttt{case\_when\_transform}$^{*}$, \texttt{in\_exists\_subquery}$^{*}$, \texttt{nullif\_coalesce}$^{*}$ \\
G8  & Snapshot / aggregation & \texttt{group\_by\_order\_limit}, \texttt{set\_snapshot\_mode}, \texttt{conditional\_aggregation}$^{*}$ \\
G9  & CEP                    & \texttt{match\_recognize}, \texttt{union\_all\_last\_value} \\
G10 & JSON / serialisation   & \texttt{json\_value\_extraction}, \texttt{json\_value\_cast}, \texttt{metadata\_virtual\_column} \\
G12 & Sink / background      & \texttt{execute\_statement\_set}, \texttt{ctas\_passthrough}, \texttt{regexp\_replace\_mask} \\
G13 & Schema / configuration & \texttt{set\_state\_ttl}, \texttt{faker\_connector\_ddl} \\
G14 & Scalar expressions     & \texttt{convert\_tz}, \texttt{timestampdiff}, \texttt{time\_series\_math}$^{\circ}$ \\
G15 & Complex / nested types$^{*}$ & \texttt{cross\_join\_unnest}$^{*}$, \texttt{nested\_row\_access}$^{*}$, \texttt{array\_functions}$^{*}$, \texttt{json\_parsing\_functions}$^{*}$ \\
\bottomrule
\end{tabular}
\caption{SQL pattern groups in FlinkEval. G11 (UDFs, extension and ML functions)
is intentionally absent: the taxonomy defines it, but its constructs are
disabled because they need registered JARs or models that the environment does
not provide; group identifiers are kept stable rather than renumbered.
$^{*}$Added in the expanded catalogue.
$^{\circ}$In the catalogue but not generated, because the environment lacks a
stable reference result. The patterns span all
three execution modes (streaming, snapshot, background/DML).}
\label{tab:pattern-groups}
\end{table*}

\subsection{Benchmark generation}

\begin{table*}[t]
\centering\small\setlength{\tabcolsep}{4pt}
\begin{tabular}{@{}p{3.2cm}p{12.2cm}@{}}
\toprule
\multicolumn{2}{@{}l}{\textbf{Entry 24706241 (water domain)}} \\
\midrule
Pattern / provenance & \texttt{temporal\_join} (G6); \texttt{suite\_ids} 4, 18, 25;
  tier~1; streaming; one repair round \\
Tables & \texttt{water\_33\_\_dim\_utility} and \texttt{water\_33\_\_dim\_region},
  both upsert tables with a primary key \\
L0 (schema-explicit) & Enrich each utility record from water\_33\_\_dim\_utility ---
  utility\_code, utility\_name and region\_code --- with the region\_name from
  water\_33\_\_dim\_region as it stood at the moment that utility record arrived,
  matching on region\_code. \\
L2.1 (business) & Can you show my utilities with the name of the region each one belongs to? \\
Reference SQL &
  \begin{minipage}[t]{12.0cm}
  \vspace{-6pt}
  {\footnotesize\begin{verbatim}
SELECT u.utility_code, u.utility_name, u.region_code,
       r.region_name
FROM `water_33__dim_utility` u
JOIN `water_33__dim_region`
  FOR SYSTEM_TIME AS OF u.`$rowtime` AS r
  ON u.region_code = r.region_code
  \end{verbatim}}
  \vspace{-10pt}
  \end{minipage} \\
Stored result & e.g.\ (U005, Cascadia Public Water, R05, Pacific Ownership Reporting) \\
\bottomrule
\end{tabular}
\caption{An example entry generated from the \texttt{temporal\_join} pattern, which
was derived from several seed pairs (\#4, \#18 and \#25). The entry stores the L0 and L2.1 requests, the reference SQL
executed on the cluster, and a captured result; L1.1--L2.4 variants are stored
in a sidecar file (Section~\ref{sec:exp2}).}
\label{tab:corpus-example}
\end{table*}

\subsubsection{Cluster-grounded generation}
\label{sec:corpus:gen}

We extract Avro/JSON schemas, including nested types, from the environment's
Schema Registry and construct Flink DDLs annotated with the descriptive
\texttt{model.*} metadata (grain, keys, changelog mode, allowed values).
Generation then proceeds per schema in five steps.

\paragraph{Quota planning.}
A planned share is assigned to ten pattern buckets, for example 18\% window
TVFs, 13\% nested types, 12\% deduplication and Top-N, 10\% interval joins,
10\% background statement sets, and 6\% pattern matching. Each bucket is split
evenly across the patterns the schema can support; the share of
unsupported patterns is redistributed. Each task is also assigned an intended
execution mode (target 45\% snapshot, 45\% streaming, 10\% background) and
zero, one or two secondary constructs (target 40 / 40 / 20\%), such as a
\texttt{CASE} derivation, a \texttt{TIMESTAMPDIFF} duration or a join with an
FK-linked table, to produce multi-pattern queries.

\paragraph{Capability-based routing.}
Each pattern declares the table capabilities it needs (an append stream with a
watermark, two such streams, an upsert dimension with a primary key, an
\texttt{ARRAY}, \texttt{ROW} or JSON column), and tasks are routed only to
tables that provide them. Routing prevents a frequent failure mode:
streaming constructs applied to updating tables.

\paragraph{Guided generation.}
An LLM (Claude Opus~5) writes one statement per task, given the routed DDLs,
up to twelve previously accepted corpus queries from the same pattern group as
examples, and a set of dialect guardrails derived from the SME audit and
the dialect gaps observed in the seed pairs. Global guardrails forbid time-relative predicates on historic data,
restrict streaming constructs to watermarked append tables, require a unique
tie-breaker in Top-N rankings, and require matching timestamp types in
interval arithmetic; pattern-specific guardrails state the required form
(e.g., \texttt{SESSION} windows are streaming-only on CC).

\paragraph{Acceptance and repair.}
Each candidate must (i)~pass a static lint that encodes the known defect classes,
(ii)~contain the target construct and match the planned execution mode,
(iii)~be accepted by the CC Flink planner (\texttt{EXPLAIN}) and execute on the
cluster, and (iv)~return at least one row on the live data. A failing candidate
is returned to the LLM with the lint violations, the Flink error, or the empty
result as feedback, up to two times. Background statements are validated by
\texttt{EXPLAIN} on each inserted query, and \texttt{CREATE TABLE \ldots AS}
statements by executing their inner query. Across the run, 375 tasks were
rejected after two repairs (137 still returned no rows, 118 still failed on
Flink, 120 still failed the lint, of which 103 lacked the target construct);
of the accepted pairs, 852 (21.8\%) required at least one repair.

\paragraph{Question generation and deduplication.}
For each accepted query, the LLM writes the L0 schema-explicit request and the
L2.1 business-language request; rows with an empty or truncated request are
discarded. Near-duplicate pairs are removed by NL and SQL similarity
(threshold 0.95), which dropped 309 of 4,212 candidates. Semantic labels are
recomputed from the SQL with the pattern detectors, and the complexity tier is
the number of distinct labels. A final pass generates the intermediate
specification levels (L1.1--L2.4) with an LLM verifier
(Section~\ref{sec:exp2}).

Execution acceptance establishes compatibility with the target environment
and a non-empty reference result; alignment of each request with its query is
produced by the LLM and is not independently validated.

\subsection{Semantic labels}
\label{sec:appendix:labels}

Table~\ref{tab:labels} presents representative labels, grouped by SQL pattern
area. Each pair carries one or more labels, so the 3,903 pairs yield 7,408
assignments over 45 observed labels (Table~\ref{tab:pattern-groups}). The five
most frequent labels account for 2,226 assignments (30.0\%):
\texttt{two\_table\_join\_agg} (597), \texttt{timestampdiff} (474),
\texttt{tumble\_tvf} (433), \texttt{convert\_tz} (396) and
\texttt{array\_functions} (326). Filters with literal rather than
clock-relative bounds are not generation targets but are still detected in 264
and 222 queries (\texttt{null\_empty\_filter},
\texttt{filter\_with\_interval\_bound}).

\begin{table}[t]
\centering
\footnotesize
\begin{tabular}{lp{4.8cm}}
\toprule
\textbf{Area} & \textbf{Labels} \\
\midrule
Window TVFs   & \texttt{tumble\_tvf}, \texttt{hop\_tvf},
                \texttt{session\_window\_tvf}, \texttt{windowed\_count\_having} \\
OVER windows  & \texttt{over\_rows\_preceding},
                \texttt{over\_range\_\allowbreak interval\_preceding},
                \texttt{over\_window\_unbounded} \\
Joins         & \texttt{temporal\_join}, \texttt{unbounded\_left\_join},
                \texttt{stream\_stream\_\allowbreak interval\_join},
                \texttt{multi\_join\_chain}, \texttt{window\_join} \\
Aggregation   & \texttt{two\_table\_join\_agg},
                \texttt{group\_by\_order\_limit} \\
Dedup/Top-N   & \texttt{dedup\_rownumber\_eq1},
                \texttt{streaming\_topn\_rownumber} \\
CEP           & \texttt{match\_recognize} \\
JSON/CDC      & \texttt{json\_value\_cast},
                \texttt{json\_value\_extraction} \\
Filtering     & \texttt{null\_empty\_filter},
                \texttt{filter\_with\_\allowbreak interval\_bound},
                \texttt{timestampdiff} \\
\bottomrule
\end{tabular}
\caption{Semantic labels grouped by Flink SQL pattern area (partial
listing of the 45 observed labels).}
\label{tab:labels}
\end{table}

\subsection{Evaluation groups}
\label{sec:appendix:groups}

Tables~\ref{tab:exp1-pattern-groups} and~\ref{tab:exp1-best} aggregate the 45 semantic labels into nine groups,
following the pattern areas of Table~\ref{tab:labels}, extended so that every
label belongs to exactly one group.
Corpus sizes (questions touching the group, over all 3,903 pairs) are given in
parentheses.

\begin{description}\footnotesize
  \item[Window TVFs (852)] \texttt{tumble\_tvf}, \texttt{hop\_tvf},
    \texttt{session\_window\_tvf}, \texttt{cumulate\_tvf},
    \texttt{windowed\_count\_having}.
  \item[OVER windows (405)] \texttt{over\_rows\_preceding},
    \texttt{over\_range\_interval\_preceding}, \texttt{over\_window\_unbounded},
    \texttt{avg\_over\_partition}, \texttt{lag\_over\_window},
    \texttt{lead\_over\_window}.
  \item[Joins (678)] \texttt{temporal\_join},
    \texttt{stream\_stream\_interval\_join}, \texttt{multi\_join\_chain},
    \texttt{window\_join}, \texttt{unbounded\_left\_join},
    \texttt{unbounded\_inner\_join}, \texttt{self\_join\_aggregation}.
  \item[Aggregation (899)] \texttt{two\_table\_join\_agg},
    \texttt{group\_by\_order\_limit}, \texttt{conditional\_aggregation}.
  \item[Dedup / Top-N (220)] \texttt{dedup\_rownumber\_eq1},
    \texttt{streaming\_topn\_rownumber}.
  \item[Pattern matching, CEP (183)] \texttt{match\_recognize}.
  \item[JSON and nested types (657)] \texttt{json\_value\_cast},
    \texttt{json\_value\_extraction}, \texttt{json\_parsing\_functions},
    \texttt{array\_functions}, \texttt{cross\_join\_unnest},
    \texttt{nested\_row\_access}.
  \item[Filters and scalar functions (1{,}903)] \texttt{null\_empty\_filter},
    \texttt{filter\_with\_interval\_bound}, \texttt{timestampdiff},
    \texttt{convert\_tz}, \texttt{regexp\_replace\_mask},
    \texttt{case\_when\_transform}, \texttt{nullif\_coalesce},
    \texttt{in\_exists\_subquery}.
  \item[Statements and set operations (605)] \texttt{execute\_statement\_set},
    \texttt{union\_all\_last\_value}, \texttt{metadata\_virtual\_column},
    \texttt{ctas\_passthrough}, \texttt{set\_state\_ttl},
    \texttt{set\_snapshot\_mode}, \texttt{faker\_connector\_ddl}.
\end{description}

\subsection{Corpus entry schema}
\label{sec:appendix:schema}

Each JSONL entry contains the following fields:
\texttt{stmt\_num} (unique integer ID),
\texttt{source\_file} (schema identifier) and \texttt{domain},
\texttt{sql\_cc} (validated Flink SQL),
\texttt{nl\_input\_generated} (schema-explicit source form),
\texttt{nl\_question\_simple} (unclarified business-language base form),
\texttt{what\_it\_exercises\_generated} (pattern description),
\texttt{labels} (list of semantic tags) and \texttt{group\_ids},
\texttt{execution\_mode} (\textsc{snapshot} / \textsc{streaming} /
\textsc{background}, the last denoting DML),
\texttt{complexity\_tier} (1, 2, or 3), and
\texttt{ddl\_refs} (list of referenced table names).
Entries additionally record the generation plan and outcome:
\texttt{target\_label}, \texttt{quota\_bucket}, \texttt{planned\_mode},
\texttt{planned\_tier}, \texttt{planned\_secondaries}, and
\texttt{generation\_repairs} (number of repair rounds, 0--2).
Paired ground-truth files (\texttt{.gt.jsonl}) store
\texttt{ground\_truth\_df} (up to 200 result rows as JSON),
\texttt{ground\_truth\_df\_truncated}, \texttt{execution\_time\_ms}, and
\texttt{sql\_execution\_error}. Difficulty sidecars
(\texttt{.difficulty.jsonl}) store the L1.1--L2.4 requests, the ranked clues,
and the verifier outcome per level.

% ===================== Appendix C =====================

\section{Evaluation Protocol and Configuration}

\label{sec:appendix:configuration}

\paragraph{Model identifiers and prompts.}
The model-comparison commands use \texttt{azure/gpt-6-astra},
\texttt{azure/gpt-5.6-sol}, \texttt{aws/claude-opus-5},
\texttt{gcp/gemini-3.6-flash}, and
\texttt{watsonx/openai/gpt-oss-120b}. Both experiments use generation prompt
version \texttt{v1}. The runner defaults are temperature 0.2 and a maximum
of 4,096 generated tokens. The judge configurations identify
\texttt{aws/claude-opus-5-5}, \texttt{azure/gpt-6-astra}, and
\texttt{gcp/gemini-3.6-flash}; the scoring adapter prefixes these identifiers
with \texttt{litellm:}. Ground-truth-anchored and face-value evaluation use
separate prompt configurations with the same model pool and role-assignment
seed. The face-value results use prompt version \texttt{v3}, which tells the
judge that the question abstracts the SQL (extra specificity is not a defect),
shows each statement's execution status (ran, failed, or not executed) as fact,
and narrows \textit{maybe} to a single unverifiable requirement. An earlier
prompt (\texttt{v1}) without these elements gave 86.8--88.0\% at L0--L1.2 and
75.7--77.7\% at L2.4--L2.1 on the same predictions; the two versions are not
comparable. The variant-generation records identify \texttt{azure/gpt-5.6-sol} as
both generator and verifier, with prompt version
\texttt{difficulty\_variants/v1}; the generation script sets temperatures
of 0.3 and 0.0 for these respective roles.

\paragraph{Prediction execution.}
The execution runner infers snapshot or streaming retrieval from the
predicted SQL. Its defaults are a 300-second submission/completion timeout,
a 1,000-row prediction-output cap, and a 60-second streaming read budget
after submission. A streaming read ends when the row cap or read deadline
is reached, or the cursor reports no further results. An empty fetch is
followed by a 0.2-second pause. The read deadline is checked between fetches,
so a blocking client call can extend the observed duration. The runner then
requests that the statement stop and waits for confirmation. Snapshot
retrieval fetches up to one row beyond the cap to detect truncation.

Execution duration covers submission and result collection and, for
streaming retrieval, the stop request. Submission or retrieval exceptions
are recorded as execution errors; an exception while stopping is logged
separately. Success therefore means no recorded error during the observation
procedure, rather than indefinite runtime correctness. Prediction-output
limits are distinct from the corpus reference-capture limits.

\subsection{Streaming execution accuracy}
We do not report execution accuracy (ExAcc), conventionally defined by
agreement between predicted and reference SQL results. Applying this metric to
streaming data requires an explicit comparison protocol for evolving inputs,
event-time and watermark alignment, and updates or retractions in output
streams. Clock-relative predicates and truncated captures further complicate
comparison. We leave the extension and validation of ExAcc to future work,
including aligned input snapshots or replay, defined observation horizons, and
comparison of evolving results.

\subsection{Confidence intervals}
\label{sec:appendix:ci}

Pass rates are proportions; Table~\ref{tab:ci} shows the 95\% confidence
interval half-width (Wilson method, $z=1.96$) for selected sample sizes at
$\hat{p}=0.5$. For $k$ passes among $n$ questions, $\hat{p}=k/n$, the interval is
\[
\frac{\hat{p}+z^2/(2n)\;\pm\;
z\sqrt{\hat{p}(1-\hat{p})/n+z^2/(4n^2)}}{1+z^2/n}.
\]
At $\hat{p}=0.5$ it is symmetric about the observed proportion; otherwise
we report its endpoints. For example, 362 passes among 394 questions
(91.9\%) yield an interval of 88.8--94.2\%. The v3 model comparison
($n=666$) has a worst-case half-width of 3.8 points.

\begin{table}[h]
\centering
\small
\begin{tabular}{rr}
\toprule
\textbf{Sample size} & \textbf{95\% CI half-width (pp)} \\
\midrule
  10 & $\pm$26.3 \\
  30 & $\pm$16.8 \\
 100 & $\pm$ 9.6 \\
 394 & $\pm$ 4.9 \\
 400 & $\pm$ 4.9 \\
 666 & $\pm$ 3.8 \\
\bottomrule
\end{tabular}
\caption{95\% confidence interval half-widths (Wilson) for pass-rate estimates
at various sample sizes, assuming $p \approx 0.5$ (worst case).}
\label{tab:ci}
\end{table}

\paragraph{Statistical scope.}
The five-model comparison uses 666 questions, limiting the resolution of
small performance differences (Table~\ref{tab:ci}). Subgroup estimates have
smaller denominators than the overall comparison, for example 28
Dedup / Top-N questions.
The judge pool overlaps with three generator families (Claude, GPT-6-Astra,
Gemini~3.6 Flash), although the Claude judge and generator use different
versions (Opus~5.5 and Opus~5). Shared model biases may favour related outputs; the adjudication
protocol does not establish the absence of this bias.
Each model and condition is evaluated in a single run.

% ===================== Appendix D =====================

\section{Additional Experimental Results}
\label{sec:appendix:experiments}

\begin{table*}[t]
\centering\small\setlength{\tabcolsep}{4pt}
\begin{tabular}{lrrrr}
\toprule
 & \textbf{$n$} & \textbf{Pass\%} & \textbf{95\% CI} & \textbf{Err.} \\
\midrule
All & 666 & 79.3 & 76.0--82.2 & 143 \\
\midrule
Window TVFs    &  57 & 91.2 & 81.1--96.2 &  3 \\
OVER windows   &  81 & 74.1 & 63.6--82.4 & 13 \\
Joins          & 121 & 71.9 & 63.3--79.1 & 17 \\
Aggregation    & 167 & 80.2 & 73.5--85.6 &  9 \\
Dedup / Top-N  &  28 & 96.4 & 82.3--99.4 &  2 \\
CEP            &  41 & 53.7 & 38.7--67.9 &  2 \\
JSON, nested   & 128 & 67.2 & 58.7--74.7 & 28 \\
Filters, scalar & 251 & 77.7 & 72.1--82.4 & 24 \\
Statements     & 132 & 78.8 & 71.1--84.9 & 95 \\
\bottomrule
\end{tabular}
\caption{GPT-6-Astra performance by pattern group on the comparison sample
(judge-pass rate with Wilson 95\% CI). Err.: statements that failed to
execute. A question with several labels contributes to every group.}
\label{tab:exp1-best}
\end{table*}

\begin{table*}[t]
\centering\small
\begin{tabular}{lrrrrrr}
\toprule
\textbf{Pattern group} & \textbf{$n$} & \textbf{GPT-6} & \textbf{Claude}
  & \textbf{GPT-5.6} & \textbf{Gemini} & \textbf{OSS-120B} \\
\midrule
Window TVFs                &  57 & \textbf{91.2} & 75.4 & 42.1 & 36.8 &  8.8 \\
OVER windows               &  81 & 74.1 & \textbf{84.0} & 79.0 & 70.4 & 44.4 \\
Joins                      & 121 & \textbf{71.9} & 69.4 & 67.8 & 63.6 & 45.5 \\
Aggregation                & 167 & \textbf{80.2} & 74.9 & 74.3 & 72.5 & 52.7 \\
Dedup / Top-N              &  28 & \textbf{96.4} & \textbf{96.4} & \textbf{96.4} & \textbf{96.4} & 64.3 \\
Pattern matching (CEP)     &  41 & 53.7 & \textbf{61.0} & 29.3 & 29.3 &  0.0 \\
JSON and nested types      & 128 & \textbf{67.2} & 63.3 & 56.2 & 51.6 & 24.2 \\
Filters, scalar functions  & 251 & 77.7 & \textbf{80.9} & 64.5 & 62.2 & 30.3 \\
Statements, set operations & 132 & 78.8 & \textbf{84.1} & 78.0 & 61.4 & 10.6 \\
\bottomrule
\end{tabular}
\caption{Judge-pass rate (\%) by pattern group for the model comparison
($n=666$; best value per row in bold). A question with several labels
contributes to every group, so rows do not sum to 666. The Window TVF group
contains only \texttt{session\_window\_tvf} and \texttt{cumulate\_tvf} here,
because the \texttt{TUMBLE} and \texttt{HOP} questions are excluded.}
\label{tab:exp1-pattern-groups}
\end{table*}

\paragraph{Query-pattern analysis.}
Tables~\ref{tab:exp1-pattern-groups} and~\ref{tab:exp1-best} identify substantial
variation across query patterns.
Dedup / Top-N (96.4\%), window TVFs (91.2\%), aggregation (80.2\%), filters
and scalar functions (77.7\%) and statements (78.8\%) yield relatively high
pass rates for GPT-6-Astra; the window-TVF group is restricted to session and
cumulate windows in this release (Table~\ref{tab:exp1-pattern-groups}).
Pattern matching with \texttt{MATCH\_RECOGNIZE} is the weakest group for
GPT-6-Astra (53.7\%, 22 of 41), GPT-5.6-Sol and Gemini~3.6 Flash (both 29.3\%),
and Claude Opus~5 reaches 61.0\%; GPT-OSS-120B passes none.
Window TVFs separate the models most: GPT-6-Astra reaches 91.2\%, Claude Opus~5
75.4\%, and GPT-5.6-Sol and Gemini~3.6 Flash only 42.1\% and 36.8\%.
Joins (71.9\% for GPT-6-Astra) and JSON and nested types (67.2\%) are the
next weakest groups. Combining joins with other patterns lowers GPT-6-Astra's
pass rate (tier~1: 30 of 43; tier~2: 32 of 38; tier~3: 25 of 40).
The temporal join label is a clear failure point: 12 of GPT-6-Astra's 16
temporal-join statements are rejected because they use \texttt{PROCTIME()},
which Confluent's Flink dialect does not support, and only 2 of 16 pass for
GPT-6-Astra, Claude Opus~5 and GPT-5.6-Sol alike (Gemini~3.6 Flash: 0).
This illustrates dialect-specific constraints that the table DDL alone does not
reveal.
Pooling the three strongest models, the lowest-performing individual labels with
at least eight questions are \texttt{temporal\_join} (12\%),
\texttt{lead\_over\_window} (28\%), \texttt{filter\_with\_interval\_bound} (47\%)
and \texttt{match\_recognize} (48\%).

Of GPT-6-Astra's 143 execution errors, all 81 are background-mode statements.
Of the remaining 62, 12 are the \texttt{PROCTIME()} errors above, 7 are
\texttt{TIMESTAMP\_LTZ} arithmetic errors, and 43 are miscellaneous parse,
planner or validation errors (for example, unsupported \texttt{MATCH\_RECOGNIZE}
clauses or unresolved functions).
Of the 62 non-background errors, 1 is nevertheless judged \textit{pass}; of the
81 background errors, 72 are.

\subsection{Per-domain results}
\label{sec:appendix:domains}

\begin{table*}[t]
\centering\small\setlength{\tabcolsep}{4pt}
\begin{tabular}{lrr@{\hspace{14pt}}lrr@{\hspace{14pt}}lrr}
\toprule
\textbf{Domain} & \textbf{Tab.} & \textbf{Pass\%}
 & \textbf{Domain} & \textbf{Tab.} & \textbf{Pass\%}
 & \textbf{Domain} & \textbf{Tab.} & \textbf{Pass\%} \\
\midrule
advertising & 7 & 78.6 & farming & 8 & 83.1 & pipeline & 7 & 81.1 \\
aid & 4 & 71.6 & forestry & 5 & 85.4 & property & 5 & 82.2 \\
aquaculture & 5 & 80.4 & homecare & 7 & 87.6 & recruitment & 7 & 83.7 \\
banking & 6 & 68.9 & hospital & 13 & 82.7 & recycling & 9 & 80.6 \\
build & 6 & 84.4 & hotel & 6 & 80.8 & renewable & 5 & 80.8 \\
charging & 7 & 85.7 & insurance & 10 & 84.7 & research & 7 & 80.4 \\
cloud & 6 & 76.7 & laboratory & 7 & 76.8 & restaurant & 7 & 75.0 \\
construction & 6 & 82.9 & livestock & 5 & 86.7 & retail & 14 & 73.9 \\
courts & 6 & 85.0 & maritime & 7 & 83.8 & rideshare & 6 & 82.1 \\
cybersecurity & 8 & 82.6 & mining & 5 & 90.0 & sports & 11 & 79.6 \\
education & 4 & 86.6 & newsroom & 6 & 83.8 & telecom & 5 & 85.3 \\
elections & 5 & 77.9 & parcel & 5 & 84.2 & transit & 7 & 87.1 \\
electricity & 4 & 77.9 & parking & 6 & 83.8 & water & 7 & 85.0 \\
entertainment & 5 & 80.5 & permits & 8 & 80.9 & weather & 4 & 74.3 \\
exchange & 6 & 80.8 & pharmacy & 6 & 81.3 & wildlife & 6 & 82.3 \\
\bottomrule
\end{tabular}
\caption{GPT-6-Astra on the full corpus of release v2 (L0, domain DDL context,
ground-truth-anchored judge), by domain. Tab.: number of tables in the
domain schema. This run has not been repeated on v3.1.}
\label{tab:domains-full}
\end{table*}

\paragraph{Per-domain variation.}
The v3.1 comparison sample provides only 11--42 questions per domain, limiting
the precision of domain-level estimates. The domain analysis below is
therefore from an earlier run on the full corpus of release v2 (same L0
question, domain DDL and judge) and was not repeated on v3.1.
Its pass rate is 81.7\% (95\% CI 80.5--82.9), compared with 79.3\% for
GPT-6-Astra on the v3.1 test sample of Table~\ref{tab:exp1}; the two runs use
different corpus releases and question sets.
(Table~\ref{tab:domains-full}, Appendix).
Domain rates range from 68.9\% (banking) to 90.0\% (mining), with a median of
82.2\% and a standard deviation of 4.3 pp. No domain retains the 100\% accuracy
observed for some small domain subsets in the sample.
The lowest pass rates occur in banking (68.9\%), aid (71.6\%), retail (73.9\%), weather
(74.3\%) and restaurant (75.0\%).
The observed heterogeneity is not statistically significant under a common-rate
model
($\chi^2=45.6$, 44 d.f., $p=0.41$), and only mining, banking and aid have a 95\%
interval that excludes the overall rate.
These results provide insufficient evidence for a systematic domain effect.
Schema size also exhibits no observed association with accuracy: domains
contain 4--14 tables, and table count is uncorrelated with pass rate (Spearman
$\rho=-0.01$).

\paragraph{Paired comparisons.}
Discordant-pair counts (first model only / second model only) for the
v3.1 comparison are: GPT-6-Astra vs.\ Claude Opus~5 57 / 59 ($p=0.93$);
GPT-6-Astra vs.\ GPT-5.6-Sol 102 / 40 ($p<10^{-6}$); GPT-6-Astra vs.\
Gemini~3.6 Flash 127 / 32; Claude Opus~5 vs.\ GPT-5.6-Sol 101 / 37; Claude
Opus~5 vs.\ Gemini~3.6 Flash 122 / 25; GPT-5.6-Sol vs.\ Gemini~3.6 Flash
74 / 41 ($p=0.003$); every frontier model vs.\ GPT-OSS-120B at least 215 / 15.
Tests are not corrected for multiple comparisons.
Adjudication was required for 483 of the 3{,}330 judgments (14.5\%), from
12.0\% (GPT-OSS-120B) to 17.9\% (GPT-6-Astra) across generators.

% ===================== Appendix E =====================

\section{Illustrative Examples}
\label{sec:appendix:examples}

\begin{table*}[t]
\centering\small\setlength{\tabcolsep}{4pt}
\begin{tabular}{@{}p{2.2cm} p{13.2cm}@{}}
\toprule
\textbf{L0: Schema-Explicit Specification} &
  Show me all active policies where the
  \texttt{policy\_end\_date} is within the next 30 days,
  including \texttt{policy\_id}, \texttt{holder\_name},
  \texttt{premium\_amount}, and \texttt{policy\_end\_date}
  from \texttt{insurance\_42\_\_dim\_policy}. \\[3pt]
\textbf{L2.1: Unclarified Underspecified Request} &
  Which of my policies are about to expire? \\[3pt]
\midrule
\textbf{Illustrative reference SQL} &
  \begin{minipage}[t]{13.0cm}
  \vspace{-6pt}
  {\small\begin{verbatim}
SELECT policy_id, holder_name, premium_amount, policy_end_date
FROM insurance_42__dim_policy
WHERE policy_end_date BETWEEN CURRENT_DATE
  AND CURRENT_DATE + INTERVAL '30' DAY
  AND status = 'ACTIVE'
  \end{verbatim}}
  \vspace{-10pt}
  \end{minipage} \\[2pt]
\textbf{Candidate SQL} &
  \begin{minipage}[t]{13.0cm}
  \vspace{-6pt}
  {\small\begin{verbatim}
SELECT * FROM insurance_42__dim_policy
WHERE policy_end_date <=
  CURRENT_TIMESTAMP + INTERVAL '30' DAY
  \end{verbatim}}
  \vspace{-10pt}
  \end{minipage} \\[3pt]
\bottomrule
\end{tabular}
\caption{Schematic insurance example of omitted query constraints.
  The candidate changes the projection, omits the active-policy and lower-date
  filters, and uses a timestamp-relative upper bound. L0 and L2.1 illustrate
  paired specification conditions; the example reports no model judgments
  or live execution results.}
\label{tab:judge-example}
\end{table*}

\subsection{Illustrative Flink SQL constructs}
\label{sec:appendix:sql-examples}

The following examples illustrate the temporal operations and data
manipulation constructs motivating FlinkEval. They are schematic examples
rather than additional benchmark measurements.

\paragraph{Windowed aggregation.}
A streaming query can count orders per customer in five-minute event-time
windows, using the window boundary columns supplied by the TVF
\cite{confluent2026windows}:
{\small
\begin{verbatim}
SELECT customer_id,
  window_start AS ws,
  COUNT(*) AS order_count
FROM TUMBLE(TABLE orders,
  DESCRIPTOR($rowtime),
  INTERVAL '5' MINUTE)
GROUP BY customer_id, window_start,
  window_end
\end{verbatim}
}

\paragraph{Temporal enrichment.}
An event-time temporal join associates an order with the exchange rate valid
at the order's event time, assuming a versioned rate table keyed by currency
and suitable event-time watermarks:
{\small
\begin{verbatim}
SELECT o.order_id, o.amount,
  r.rate, o.amount * r.rate AS usd
FROM orders AS o
JOIN exchange_rates FOR SYSTEM_TIME
  AS OF o.$rowtime AS r
ON o.currency = r.currency
\end{verbatim}
}

\paragraph{Background data manipulation.}
Background statements materialise query results in sink tables backed by
Kafka topics:
{\small
\begin{verbatim}
INSERT INTO `customer_summary`
SELECT customer_id,
  COUNT(*) AS total_orders,
  SUM(amount) AS total_spend
FROM `orders`
GROUP BY customer_id
\end{verbatim}
}
Confluent's \texttt{ON CONFLICT} extension additionally supports deduplication:
{\small
\begin{verbatim}
UPSERT INTO `loyalty_scores`
SELECT customer_id, score
FROM enriched_events
ON CONFLICT DO DEDUPLICATE
\end{verbatim}
}

\paragraph{Potential clarification interaction.}
The following dialogue illustrates a possible extension to interactive
clarification; the current experiments use appended oracle clues.
\begin{quote}\small
\textbf{User:} Which of my recent ad impressions were actually viewable?\\
\textbf{System:} I can help with that.
  Which table contains your impression data, and over what time window
  should ``recent'' be defined?\\
\textbf{User:} The \texttt{fact\_impression} table, last 2 hours.\\
\textbf{System:} \texttt{SELECT * FROM `fact\_impression` WHERE served\_at
  > CURRENT\_TIMESTAMP - INTERVAL '2' HOUR AND viewable = TRUE}
\end{quote}

\subsection{Illustrative analysis of omitted constraints}
\label{sec:appendix:judge-example}

Table~\ref{tab:judge-example} presents a schematic insurance example of
how question specification affects SQL evaluation. The requests and SQL
are illustrative, rather than a traced corpus entry or recorded model
prediction; no empirical judge verdicts are associated with this example.
The two requests illustrate the L0 and L2.1 specification conditions.

The candidate differs from the illustrative reference SQL in four respects:
(1)~\texttt{SELECT *} instead of the four explicitly requested columns;
(2)~the \texttt{status = 'ACTIVE'} predicate is absent, so cancelled and
lapsed policies can also be returned;
(3)~the lower bound \texttt{policy\_end\_date >= CURRENT\_DATE} is absent,
allowing already-expired policies; and
(4)~the upper bound is expressed relative to \texttt{CURRENT\_TIMESTAMP}
rather than \texttt{CURRENT\_DATE}, changing the temporal expression.
The effect of the last difference depends on the column type and the engine's
date/timestamp coercion rules; it does not by itself establish a type error.

Under reference-anchored evaluation, the reference supplies the projection,
active-policy filter, and bounded 30-day horizon. Under face-value evaluation,
L0 explicitly supplies these requirements, whereas L2.1 leaves the projection,
status restriction, and precise horizon unstated. Returning already-expired
policies remains problematic even for ``about to expire'': the candidate has
no lower date bound. For example, an active policy that ended yesterday
satisfies the candidate's upper-bound condition but is excluded by the
reference. This logical counterexample illustrates a semantic discrepancy
without relying on a particular result capture or judge score.

% -------------------------------------------------------

\end{document}